# nanoNET: Machine Learning Platform for Predicting Nanoparticles Distribution in a Polymer Matrix

Kumar Ayush, Abhishek Seth and Tarak K Patra[*]

Department of Chemical Engineering, Center for Atomistic Modeling and Materials Design and Center for Carbon Capture Utilization and Storage, Indian Institute of Technology Madras, Chennai TN 600036, India

**Abstract:**

Polymer nanocomposites (PNCs) offer a broad range of thermophysical properties that are linked to their compositions. However, it is challenging to establish a universal composition-property relation of PNCs due to their enormous composition and chemical space. Here, we address this problem and develop a new method to model the composition-microstructure relation of a PNC through an intelligent machine learning pipeline named nanoNET. The nanoNET is a nanoparticles (NPs) distribution predictor, built upon computer vision and image recognition concepts. It integrates unsupervised deep learning and regression in a fully automated pipeline. We conduct coarse-grained molecular dynamics simulations of PNCs and utilize the data to establish and validate the nanoNET. Within this framework, a random forest regression model predicts the NPs' distribution in a PNC in a latent space. Subsequently, a convolutional neural network-based decoder converts the latent space representation to the actual radial distribution function (RDF) of NPs in the given PNC. The nanoNET predicts NPs distribution in many unknown PNCs very accurately. This method is very generic and can accelerate the design, discovery, and fundamental understanding of composition-microstructure relations of PNCs and other molecular systems.

**Keywords:** Deep Learning, polymer nanocomposites, AI, Autoencoder, Radial Distribution Function, Molecular Simulation, Random Forest, Feature Engineering

[*]Corresponding author, E-mail: tpatra@iitm.ac.in

# INTRODUCTION

Polymer nanocomposites (PNC) offer a broad range of properties that are intricately connected to the spatial distribution of nanoparticles (NPs) in the polymer matrices. Understanding and controlling the distribution NPs into a polymer matrix is a significantly challenging task.[1–4] The structure, morphology, and phase behavior of PNCs are determined by the complex interplay between the packing entropy of particles with varying size and shape and effective interparticle interaction with varying strength and range.[4–11] A wide range of NPs distribution is possible in a polymer matrix depending on various factors, including NP-polymer interaction, NP size, NP shape, NP surface treatment, and NP concentration. Moreover, the spatial distribution of NPs in a polymer matrix is appeared to be very sensitive to these composition parameters. For example, NPs tend to aggregate for weak NP-polymer interaction due to the depletion force.[12–14] On the other hand, stronger NP-polymer interactions lead to steric stabilization, dispersion, and bridging of NPs in a polymer matrix.[15–18] Often, dispersion to agglomeration transitions occur as a function of nanoparticle concentration in a polymer matrix.[19] The extent of dispersion and aggregation is influenced by the details of the composition of a PNC. Therefore, controlling the thermophysical properties of a PNC requires a deeper understanding of composition-microstructure correlations. Molecular simulation and liquid state theories are extensively used to establish such correlations.[15,16,20–22] However, a comprehensive understanding is still lacking, and no generalized theory can predict the long-range spatial distributions of NPs in a polymer matrix for the broad range of composition parameters. A significant challenge in establishing such composition-microstructure relations in PNCs is their infinitely large phase space due to the enormous range of materials selection, surface chemistries, NP-polymer size ratio, NP concentration, and processing methods. Exploring the entire range of these parameters is not feasible experimentally or computationally.

We aim to address this challenge via machine learning. Recently, machine learning has shown tremendous success in solving a wide range of materials problems, such as identifying the local structure of metallic nanoparticles on an oxide-support and defective graphene sheets,[23] automated recognition of metal nanoparticles deposited on pyrolytic graphite,[24] atomic structure classification and prediction,[25–27] nanostructure identification from X-ray, SEM (scanning electron microscopy) and TEM (transmission electron microscopy), SAS(small-angle scattering) data,[28–32] and classification and prediction of sequence-defined morphologies of copolymers.[33–35] Machine learning methods are also recently used to predict the properties of PNCs,[36–39] such as dielectric constant, rubbery modulus, and glassy modulus. However, developing a machine learning workflow that predicts the long-range spatial correlation of NPs based on the composition of a PNC remains elusive. The primary bottleneck in building such a model is the disparity in dimensions of input and output. The unusual feature of this problem is that the input vector's size is very small in comparison to that of the output vector. The input vector represents the composition parameters such as NP concentration, NP-polymer interaction, NP-NP interaction, size of a NP and polymer chain length of a PNC. The output vector represents local concentrations of NPs at different interparticle spacing in the PNC. Usually, the size of the input vector is about five, while that of the output vector is about a hundred. Therefore, approximating the function $f: X \rightarrow Y$, where X and Y are the composition vector and radial distribution function (RDF), respectively, is challenging.

We attempt to solve the above problem using efficient encoding and decoding techniques that are used in computer vision and image recognition. An efficient encoding and decoding of data are critical for the performance of ML models. In ML literature, the encoding of data is commonly known as feature learning.[40] It extracts the unique characteristics from data and represents it in a lower dimensional continuous space representation. In the context of materials design and structure-property modeling, several feature learning techniques are used to represent the vast chemical space of molecules in machine-readable numerical form.[41] These features serve as an input to a machine learning regression model that predicts the properties of molecules and materials.[42] An appropriate feature selection technique is decided depending on the nature of data and target correlations. One-hot encoding, chemical tree, motif-based fingerprints, property coloring, and autoencoding are a few techniques that have been recently developed and employed in materials problems.[35,43–46] There feature learning techniques are commonly applied to the input data, which is typical vast in size. On the contrary, the current problem deals with high dimensional output and demands feature extraction from RDF for building an efficient ML model. Here, for the first time, we employ feature engineering on spatially correlated data, viz., RDF of NPs in a polymer matrix. In particular, we use an unsupervised dimensionality reduction technique to encode an RDF to a lower dimensional latent space and subsequently decode it back to the RDF space. A convolutional neural network autoencoder is used for this purpose.[47] We choose an appropriate architecture of the autoencoder that creates a latent space of the desired dimension, which is comparable to the dimension of the composition vector. A supervised regression technique is then used to build a function that correlates the composition vector to the latent space vector, as they are of comparable dimensions. This way, the regression model can predict the latent space vector of a given composition vector, which can be converted to RDF by the decoder part of the autoencoder. We envision that this combination of supervised and unsupervised learning will tackle the problem of predicting the radial distribution of NPs in a polymer matrix based on composition information.

Therefore, the objective is to establish an automated ML framework that could take the composition parameters as input and predict the long-range radial distribution of NPs in the polymer matrix. Towards this end, we consider a phenomenological model of PNC and perform coarse-grained molecular dynamics (CGMD) simulations within a carefully selected range of compositional parameters, which span a wide region of the phase space, including experimentally known phases. We then quantify the aggregation and dispersion states of filler NPs from the CGMD simulations data. In particular, we calculate the long-range radial distribution function (RDF) for 80 different combinations of composition parameters. This data is utilized to build the machine learning framework, code name nanoNET, that elucidates the composition-structure relationships. The nanoNET provides a robust machine learning method that enables intelligent interpolation in the composition and chemical space of a PNC without directly measuring its properties via physics-based computer simulations or experiments. This machine learning approach can accelerate the characterization, prediction, and fundamental understanding of polymer nanocomposites and other molecular systems.

**MODEL AND METHODOLOGY**

**Model.** A very generic phenomenological coarse-grained (CG) model of PNC is considered for this study. Here polymer chains are represented as a coarse-grained bead-spring model of Kremer and Grest[48] wherein a pair of monomers interact via the Lennard-Jones (LJ) potential of the form $V(r) = 4\varepsilon\left[\left(\frac{\sigma}{r}\right)^{12} - \left(\frac{\sigma}{r}\right)^{6}\right]$. The $\epsilon$ is the unit of pair interaction energy and the size of a monomer (*d*) is σ. In addition, two adjacent coarse-grained monomers of a polymer chain are connected by the Finitely Extensible Nonlinear Elastic (FENE) potential of the form $E = -\frac{1}{2}KR_0^2 \ln\left[1 - \left(\frac{r}{R_0}\right)^2\right]$, where $K = 30\epsilon/\sigma^2$ and $R_0 = 1.5\sigma$ for bond length $r \leq R_0$ and $E = \infty$ for $r > R_0$. The LJ interaction between a pair of monomers is truncated and shifted to zero at a cut-off distance $r_c = 2.5\sigma$. The NP is also modeled via the LJ potential. The diameter of an NP (*D*) is varied from $2\sigma$ to $5\sigma$. The NP-NP interaction is truncated and shifted to zero at a distance $r_c = 2.5D$ to represent attraction between them. The NP-NP interaction ($\epsilon_{NP-NP}$) is varied from $0.1\epsilon$ to $2.5\epsilon$. The polymer-NP ($\epsilon_{P-NP}$) interaction is varied from $0.1\epsilon$ to $1.5\epsilon$. The polymer-NP interaction is truncated and shifted to zero at a cut-off distance $r_c = 2.5 \times (D + d)/2$. The chain length of the polymer is varied between *N=25* to *40*.

**MD Simulations and Data Curation.** We conduct CGMD simulations of the PNC for a wide range of polymer-NP interaction, NP concentration, NP-NP interaction, NP-polymer size ratio; and generate a database for training and developing the machine learning framework. The CGMD simulations are performed in an isothermal isobaric ensemble (NPT), wherein the temperature and pressure are controlled by a Nose-Hoover thermostat and barostat, respectively. We use the velocity-Verlet algorithm with a timestep of $0.005\tau$ to integrate the equation of motion. Here, $\tau = \sigma\sqrt{\frac{m}{\epsilon}}$ is the unit of time, and *m* is the mass of a monomer, All the simulations are conducted for a reduced temperature $T^* = Tk_B/\epsilon = 1$, and a reduced pressure $P^* = P\sigma^3/\epsilon = 1$, using the LAMMPS open-sourced code.[49,50] All the systems are equilibrated for $10^7$ MD steps, followed by a production run of $10^7$ MD steps. During these production MD simulations, we collect 10000 configurations to calculate the RDF of a system. We carry out 80 MD simulations each for a distinct combination of *D*, *N*, *ρ*, $\epsilon_{NP-NP}$, and $\epsilon_{P-NP}$. The concentration of NPs (*ρ*) is calculated as the ratio of the number of NPs and the equilibrium simulation box volume. Combinations of all these five parameters provide a wide range of RDF, including microscopic phase separation, steric stabilization, and bridging of NPs via polymer layers. We calculate RDF within a distance of $10\sigma$. We consider a bin width of $0.1\sigma$ for the RDF calculation. Therefore, an RDF is a one-dimensional array of length 100. We plot an RDF on a cartesian coordinate for each set of composition parameters. The magnitude of an RDF in our data set varies from 0 to 300. We change the ordinate of the plot to a logarithmic scale to display a wider range of variation more compactly. We fix axes of the plots from 0 and 10 for abscissa, whereas from 0.001 and 300 for ordinate for uniformity. The RDF plot is rendered as a greyscale image, wherein the RDF is a white line with a black background. This procedure is followed to prepare 80 greyscale images, each corresponding to a specific set of five composition parameters.

**Convolutional Neural Network Autoencoder.** A convolutional neural network (CNN) is a special class of multi-layer artificial neural networks that is capable of recognizing visual patterns automatically and adaptively through learning.[51] The architecture of CNNs is inspired by the mechanism of the optical system of living creatures. It very accurately retains the correlated information between the pixels of an image, and, thus, is very efficient in feature extraction from images. CNNs have shown tremendous success in computer vision and image classification problems.[52,53] We employ CNN to extract features from RDF images and create a unique latent space representation of RDF by constructing an autoencoder. The CNN autoencoder has two main parts - an encoder that maps the input image to a smaller size vector and a decoder that reconstructs the image from its encoded vector. A schematic representation of the CNN autoencoder, which is used in the present study, is shown in Figure 1. An RDF image, which is calculated from an MD trajectory, is fed to the CNN through its input layer. It produces an identical RDF image in its output layer. The output RDF image corresponds to a morphology, which closely resembles the input one. Each RDF image is a grayscale image, which can be represented as a three-dimensional (3D) array with a depth of 1. The array rows correspond to the x-coordinate of the image, and the columns correspond to the y-coordinate. The array contains at each coordinate pair (x and y) a value, typically an integer between 0 and 255, specifying the level of greyness. The width and height of the images are 64 pixels each. Therefore, each RDF image is converted to a tensor of size 64x64x1. A total of 60 RDF images are repeatedly fed to the CNN autoencoder during the training.

The architecture of the encoder consists of six convolution layers and six pooling layers arranged alternatively. The convolution layers perform the convolution of image segments and their activation. During the convolution, a kernel, which is a small matrix of numbers, is applied to the input tensor. A dot product between the kernel and the input tensor is calculated at each location of the tensor and passes as the output value in the corresponding position of the output tensor. We use a matrix of size (3,3) for the kernel of the convolution operation. The kernel moves across the input tensor with a stride of unity. Here the stride is defined as the distance between two successive kernel positions. This operation does not allow the center of the kernel to overlap with the outermost elements of the input tensor. Padding is used to overcome this situation wherein rows or columns of zeroes are added to each side of the input tensor. This ensures that the center of the kernel can fit on all the extreme elements of the input tensor. These operations produce a planer feature map. We repeat the entire process multiple times with different kernel matrices to create a 3D feature map of the input. Therefore, the output of a convolution layer will be a tensor of length and width equal to that of the input. The width of the tensor will depend on the number of filters, which is the number of planner feature maps generated during the convolution operation. The number of filters in the successive convolution layers of the CNN are 64, 32, 32, 16, 16, and 8, respectively. The kernel parameters are learned during the training. The output of the convolution layer is also passed through an activation function. Subsequently, a pooling layer, which is placed in between two successive convolution layers, shrinks the in-plane dimension of its input tensor. The Max-pooling method is used for the pooling operations.[54] It extracts patches (subregions) from the input tensor and outputs the maximum values of the patches while discarding all other numbers. Therefore, the features are gradually reduced to lower dimensional representation as the network grows via successive convolution and pooling operations. The pooling operation changes the in-plane dimension, while the convolution operation changes the depth of the tensor. This architecture of the encoder creates a latent space of dimension 1x1x8 for an input of dimension 64x64x1. Decoding of the latent space tensor to the original RDF image is done by the decoder part of the network. It deconvolutes the latent space tensor to the actual RDF image. Usually, the decoder's topology is the encoder's mirror image. We chose twelve hidden layers for the decoder part of the network, six of them are transposed convolutional layers, and the remaining six are upsampling layers. A transposed convolution layer, which is also known as a deconvolution layer, increases the depth of the input matrix. The filters in the successive deconvolution layers are 8, 16, 16, 32, 32, and 64, respectively. A kernel of size (3,3) is used for the deconvolution operations. An activation function activates the output of the deconvolution layer. An upsampling layer, which is placed between two successive deconvolution layers, yields a feature map with an in-plane dimension greater than its input. These sequential deconvolutions and upsampling operations convert the latent space tensor of dimension 1x1x8 to the output tensor of dimension 64x64x1.

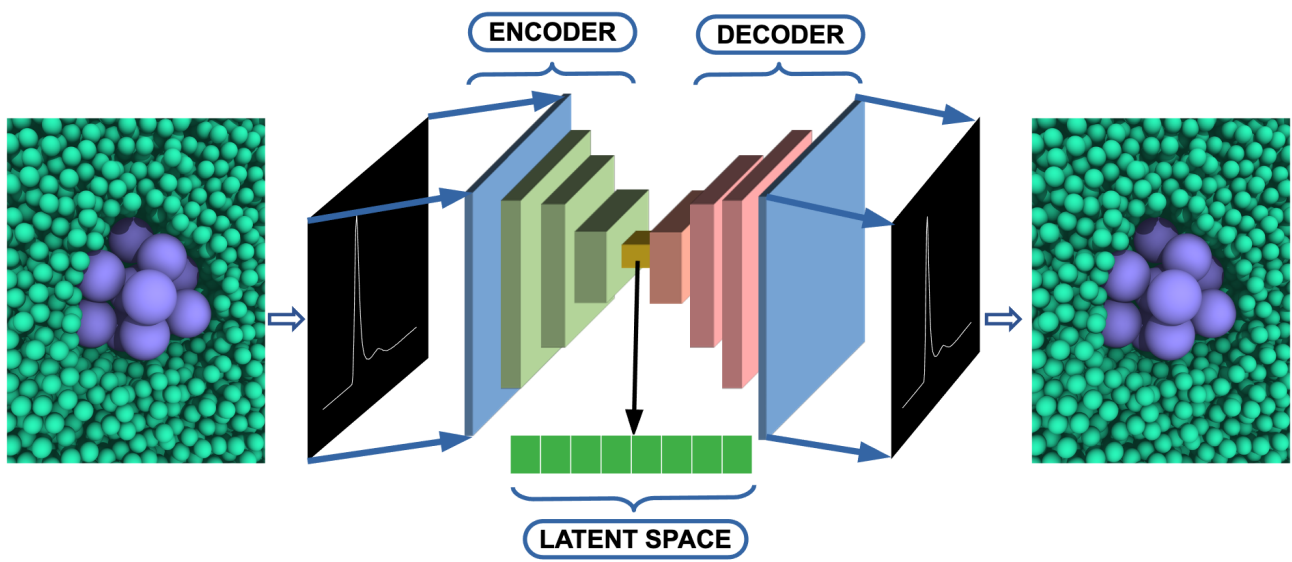

*Figure 1: A schematic representation of the convolution neural network autoencoder that encodes the image of an RDF to a lower-dimensional latent space. RDFs are calculated from MD trajectories and converted into grayscale images and feed to the autoencoder. The autoencoder reconstructs the RDF images at the output layer, which closely resembles the input layer RDF and the microstructure.*

This autoencoder links the convolutional layers, pooling layers, de-convolution layers and upsampling layers without any intermediate dense neural network. Such an autoencoder topology ensures faster training convergency and attains a higher quality local optimum. The number of layers, dimension of the input image, number of filters, stride and padding are decided based on our initial studies to improve the accuracy and efficiency of the CNN. This network shows better performance for our system than any other state-of-the-art method. The mathematical details underlying the construction of the CNN autoencoder can be seen elsewhere.[55,56] We use rectified linear unit (ReLU) function[57,58] as the activation function for the convolution and deconvolution layers. A standard backpropagation algorithm is used for the network training. A gradient-based stochastic optimization algorithm viz. Adam optimizer is used to optimize the parameters of kernels during the backpropagation.[59] We use Keras,[60] an open-sourced application programming interface (API), for constructing the autoencoder.

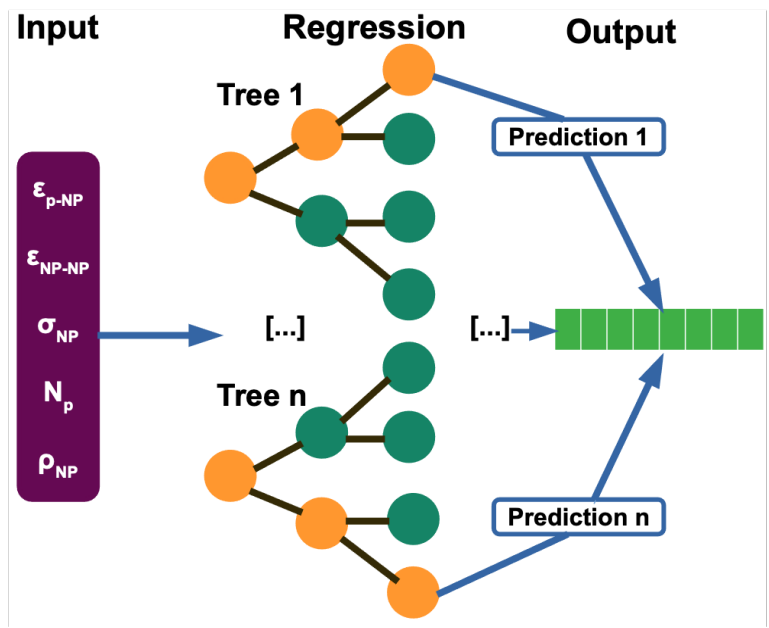

*Figure 2: A schematic representation of a random forest that takes system variables viz., PNC composition parameters as input and predicts latent space representation of the corresponding NPs' RDF in the PNC.*

**Random Forest:** The Random Forest (RF) algorithm is an ensemble learning method that can perform regression using a decision tree framework.[61,62] Ensemble learning is a process that gives a final prediction by combining the forecasts from multiple models trained over the same data. It uses bootstrap and aggregation, commonly known as bagging, which involves averaging predictions from several decision trees to give the final output.

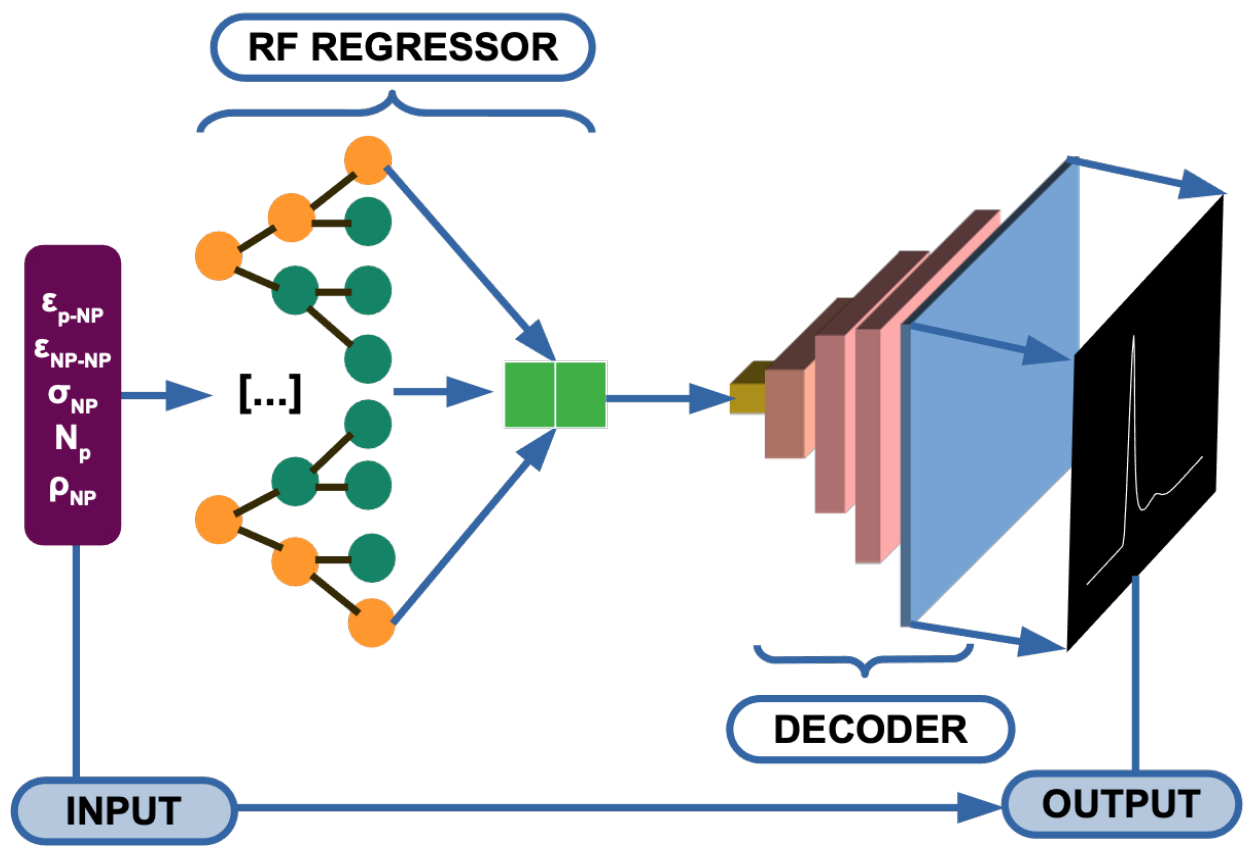

*Figure 3: nanoNET. The random forest regressor and the decoder part of the convolutional neural network autoencoder is integrated to predict the RDF of NPs in a polymer matrix. The input to this machine learning tool is the composition parameters of a PNC viz., NP-NP interaction, NP-polymer interaction, polymer chain length, NP diameter, NP concentration, and the output is its long-range RDF of NPs.*

A schematic representation of a random forest that receives composition parameters of a PNC as inputs and predicts latent space representation of the corresponding RDF is shown in Figure 2. The main parameters in this model are the number of trees and the size of the random subsets of features to consider when splitting a node. We systematically study the performance of RF for varying number of trees and select the optimal number that gives minimum error in prediction. A default value of one is used for the random subset of features while splitting a node. We build the RF regression model using scikit-learn.[63]

**nanoNET Workflow.** The success of LeNET,[64–66] and its variants such as AlexNet,[52] VGGNet,[67] GoogLeNet,[68] ResNet,[69] and DenseNet[70] in hand-written digits and image recognitions has revolutionized the field of computer vision and promoted the development of deep learning tools across disciplines. These deep learning models are primarily backpropagation convolutional neural networks that extract meaningful low-dimensional features of an image. They have multiple processing layers to learn images with multiple levels of abstraction. We adopt a similar approach to extract the spatial correlation of NPs that are distributed in a polymer matrix and link it with the composition of the PNC. The machine learning framework for predicting the long-range NPs distribution in a polymer matrix, which is abbreviated as nanoNET, has two main components – a regressor and a decoder. As discussed in the previous section, the decoder is a deconvolution neural network, which is an integral part of the CNN autoencoder. We use an RF for the regression task, as discussed in the previous section. The regressor and the decoder are connected, as shown in Figure 3 to predict the RDF of NPs in a polymer matrix. Both the RF and CNN autoencoder are trained previously with the available information (training data). For any given composition, the regressor predicts the latent space representation of the NPs' RDF, and the decoder converts it to the real 2D space representation of an RDF. Within this predictive framework, the composition matrix, which is the input vector, has a dimension of five. They represent the size of NP (*D*), polymer chain length (*N*), NP concentration (ρ), NP-NP interaction ($\epsilon_{NP-NP}$), and polymer-NP interaction ($\epsilon_{P-NP}$). The dimension of the latent space is 8. The output dimension is 64x64, which corresponds to an RDF image. The size of the latent space vector and output matrix is decided based on initial trials to improve the accuracy and efficiency of the ML pipeline. The performance of nanoNET as a function of the latent space dimension is discussed in the supporting information (SI).

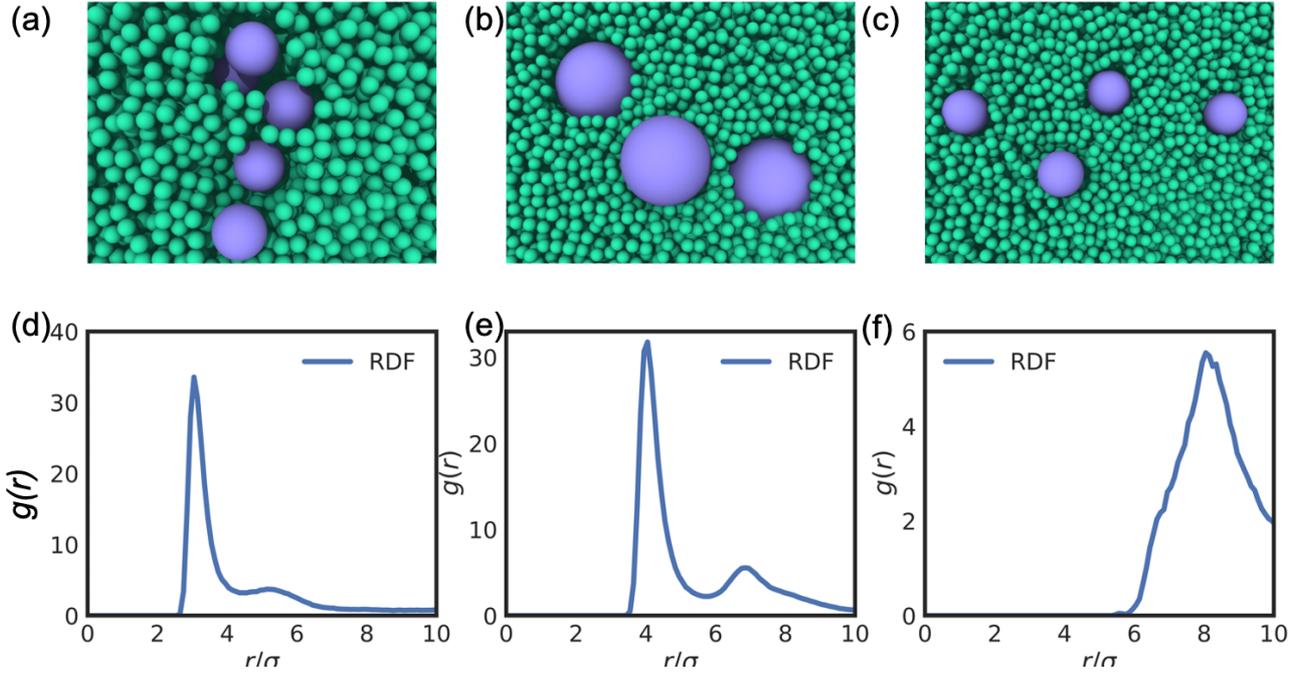

*Figure 4: MD snapshots and RDFs of NPs in a polymer matrix are shown for three representative cases. The (a), (b) and (c) correspond to composition parameters ( $D=3\sigma$, $N=40$, $\rho=0.00072$, $\epsilon_{NP-NP}=0.1\varepsilon$ and $\epsilon_{P-NP}=0.1\varepsilon$ ), ($D=3\sigma$, $N=40$, $\rho=0.00032$, $\epsilon_{NP-NP}=0.5\varepsilon$, and $\epsilon_{P-NP}=1.0\varepsilon$) and ($D=3\sigma$, $N=40$, $\rho=0.00082$, $\epsilon_{NP-NP}=0.2\varepsilon$, and $=0.4\varepsilon$), respectively. The green and blue beads are monomers of the polymers, and nanoparticles, respectively. The (d), (e) and (f) are the respective RDFs.*

**RESULTS AND DISCUSSION**

The nanoNET is built using a CNN autoencoder and an RF regressor. We use 68 RDFs to build these two components of the nanoNET. The performance of the nanoNET is tested for 12 RDFs that are not used for model building. These 80 RDFs are calculated for a selected range of composition parameters that yield a large variety of NP distributions in a polymer matrix. Figure 4 shows a few representative cases of NP distributions. We observe microscopic phase separation for a weak NP-polymer interaction wherein the first RDF peak occurs at $r \approx 3\sigma = D$. This corresponds to a complete unmixing of NPs and polymers. A layer of polymer is present in Figure 4b that defines the bridging of NPs, and the corresponding peak position is seen for an NP-NP separation distance of $r \approx 4\sigma = D + d$. Figure 4c represents a case when nanoparticles are well dispersed in the polymer matrix. The corresponding RDF shows a very low pick height, and the peak position is around a separation distance of $8\sigma$ for the case of *D=3σ*. This is a signature of a weak effective potential of mean force between NPs. Overall, our CGMD simulation-identified phases of PNC that are commonly seen in experiments,[71–74] and our machine learning protocol is targeted to establish composition-microstructure relations of PNCs that are experimentally realizable.

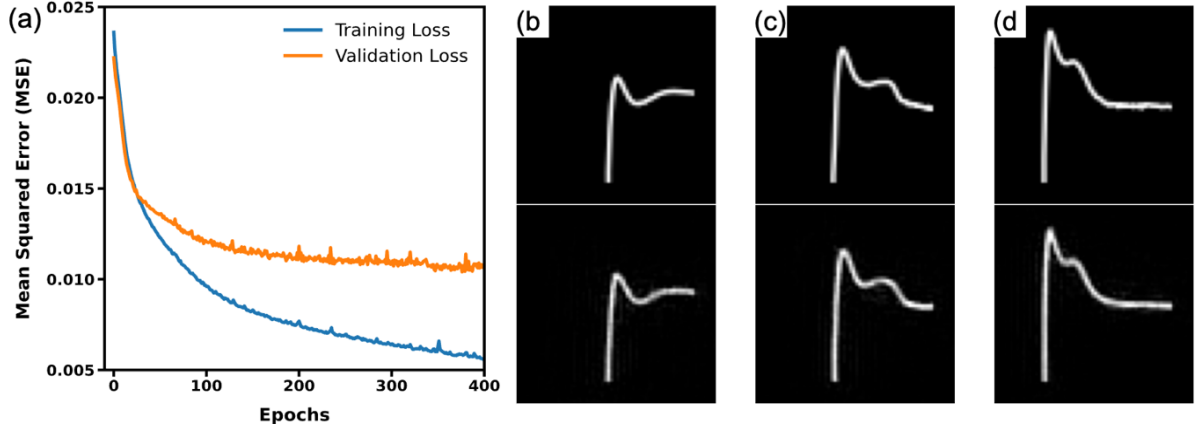

*Figure 5: Training and validation of the convolutional neural network autoencoder. Sixty-eight RDFs are used to build the convolutional autoencoder. The loss function i.e., MSE during the training is shown for training and validation data sets in (a). Three representative input RDF images and corresponding output RDF images are shown in (b), (c) and (e). The top and bottom panels in (b), (c) and (d) correspond to the actual input images and predicted output images, respectively.*

First, we focus on CNN autoencoder training with the RDFs data. All RDFs are converted to grayscale images of dimension 64x64 for machine learning purposes as discussed in the method section. We create two data sets – training and validation for establishing the autoencoder model. The training validation sets consist of 80% and 20% of the total RDF data. The CNN autoencoder is trained for 400 epochs. The mean square error during the training drops rapidly in the early stage of the training, as shown in Figure 5a for both the training and validation sets. The reconstruction loss, which is defined as the mean absolute error (MSE) in prediction, after the training is found to be ~0.005 for the training data set. The autoencoder compresses a two-dimensional image of size 64x64 to a one-dimensional array of size 8. We decide the dimension of the latent space based on many initial studies. Our initial studies suggest that latent space of size 8 most accurately predicts the RDF. Details of these analyses can be seen in the supporting information (SI). Once the training completes, we evaluate the performance of the autoencoder. The input image and output image of a trained CNN autoencoder are compared in Figures 5b, c, and d, for three representative cases. The visual inspection suggests close agreement between input and output images. The MSE in prediction is below 2%. Therefore, we infer that the CNN autoencoder creates a unique latent space representation of the RDFs.

Now, we build an RF model to predict the latent space vector of an RDF based on the information of the composition of a PNC. Each feature and target are scaled individually such that they vary in the range between zero and one. The normalized dataset is split randomly into two subsets- one for training the RF regressor, and another one for testing the performance of the regressor. The training and test subsets contain 80% and 20% of the total data, respectively. We vary the number of decision trees in the RF to identify the best model. Figure 6a shows the MSE as a function of decision trees in the RF. It indicates that the best performance of the RF regressor can be achieved for over 20 decision trees. The performance of an RF with 20 decision tresses is compared in Figure 6b. The predicted and actual latent space values are in close agreement. The coefficient of determination ($r^2$) is 0.98 and 0.84 for training and test sets, respectively. We infer that this RF model can serve as a surrogate model for predicting the latent space vector of an RDF image. We note that one can build an RF model that directly predicts the RDF for a given composition vector of a PNC. In that case, an RDF is a one-dimensional array of length 100, each representing the local density of NPs within a range of 0 to 10σ NP-NP separation distance. As shown in **SI,** a RF regressor can't establish such a correlation between two vectors of disparate dimensions. Also, predicting 100 target values correctly from the 5 features/inputs is equally error-prone for other AI/ML algorithms like XG Boost, deep neural network, and ridge regression. Therefore, instead of directly predicting an RDF, we build a regression model that predicts a lower dimensional representation of the RDF.

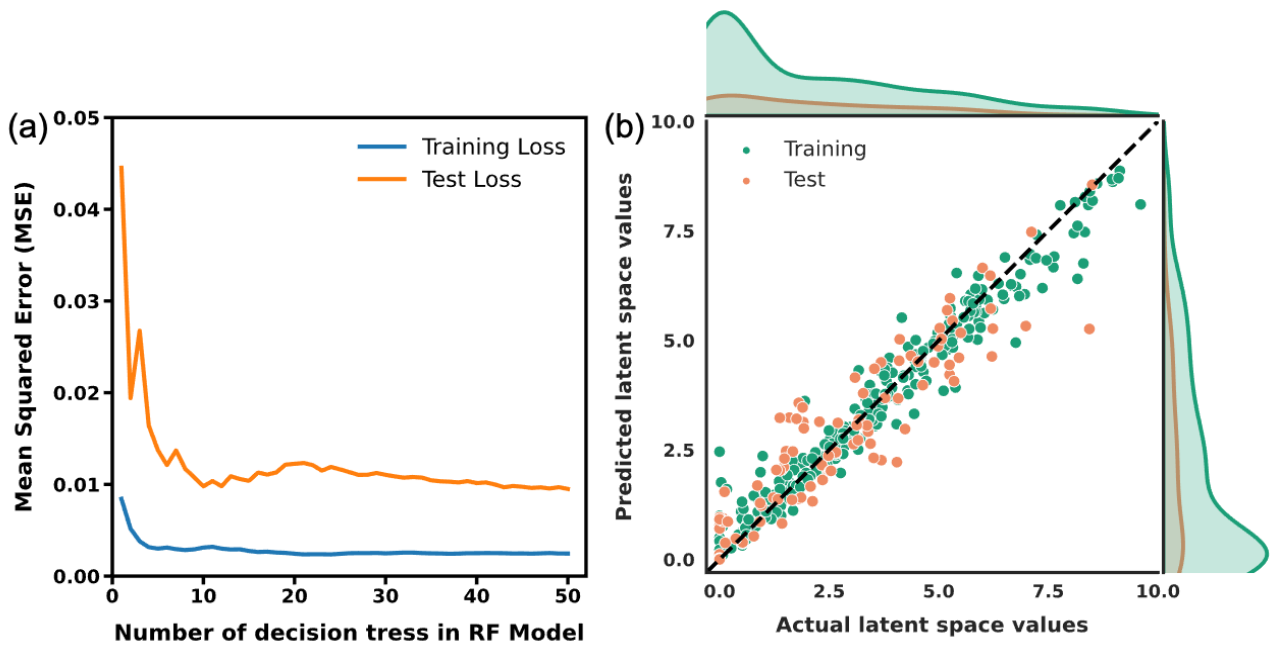

*Figure 6: Development of a random forest (RF) model and its performance evaluation. Mean square error is plotted as a function of the number of decision tress in an RF in (a). The MSE is appeared to reach a plateau when the number of decision tress is around 20. The actual and predicted latent space variables are compared in (b) for training and test sets.*

We now integrate the RF model and the decoder part of the CNN autoencoder and build the nanoNET that predicts the RDF of NPs for a given PNC, as schematically shown in Figure 3. The nanoNET receives composition information of a PNC viz, NP-polymer interaction strength ($\epsilon_{P-NP}$), NP-NP interaction strength ($\epsilon_{NP-NP}$), NP concentration ($\rho$), NP diameter ($D$) and length of a polymer chain ($N$) as input. The RF model predicts the latent space vector for this given composition vector. The predicted latent space vector is fed to the decoder. The decoder constructs the RDF of NPs in the system and produces as an output of the nanoNET. The predicted RDFs from the nanoNET are compared to the actual RDFs that are computed using CGMD simulations in Figure 7 for twelve representative cases. The set of composition parameters for all the representative cases is tabulated in SI. The predicted first peak height and position of an RDF for all the cases are in very close agreement with their actual value. The model is able to predict a wide extent of structural correlations from a very weak correlation in Figure 7k, where the peak height is about 4 to a very strong correlation in Figure 7l, where the peak height is about 100. Similarly, the first peak position shifts from around 2.5σ ( Figure 7b, 7e, 7g, 7k) to around 5σ (Figure 7j). The model makes a very accurate prediction for the entire range of the first coordination shell for all the cases. For a few cases (Fig 7a, b, and c), the second peak height and location are slightly overpredicted. Overall, the shape of the predicted RDF curve is largely in agreement with that of the actual curve for all the cases. Moreover, the phase behavior of the PNC is predicted very accurately. For example, predicted microscopic phase separation (Figure 7l), bridging of nanoparticles by a polymer layer (Figure 7f) and dispersion (Figure 7k) are well in agreement with the actual phases.

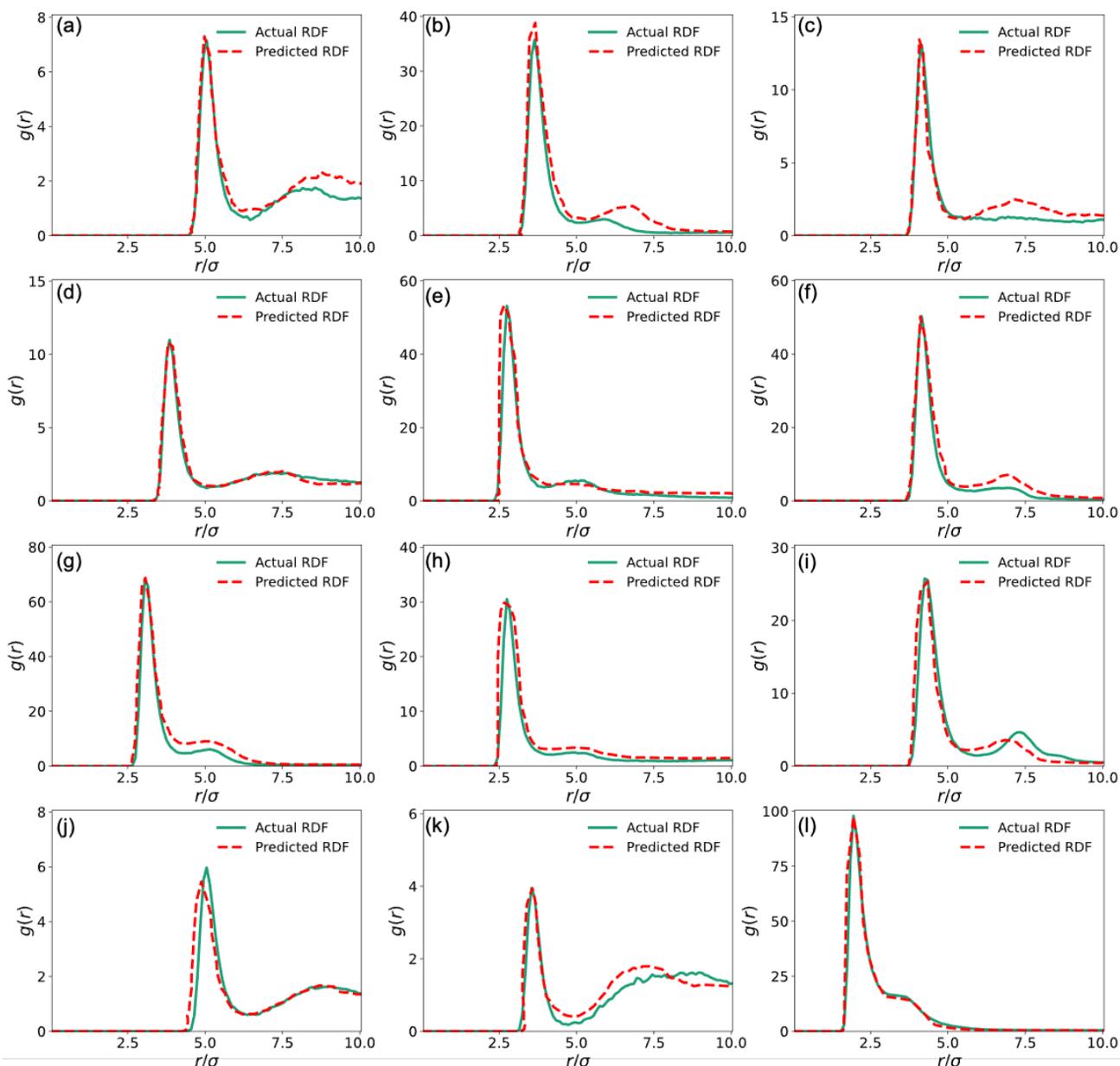

*Figure 7: Performance of nanoNET. The actual RDF of nanoparticles in a polymer matrix is compared with that of the predicted one for twelve representative cases. Each case corresponds to a distinct combination of composition parameters. The composition parameter value for each of these cases can be seen in SI.*

**CONCLUSIONS**

AI and ML offer exceptional promises to predict the behavior of a vast chemical and composition space of a material based on a relatively smaller number of training data points. However, it is challenging to develop an ML workflow that correlates two vectors of disparate dimensions. Here we tackle this problem using feature engineering for polymer nanocomposites. Specifically, we develop a model that correlates the composition of a polymer nanocomposite and the long-range spatial distribution of nanoparticles in it. As the composition vector and radial distribution function are of disparate dimensions, we use a data encoding technique to solve this problem. The quality of data encoding is very critical for the success of any ML pipeline and model development. We show that an unsupervised convolutional neural network provides a unique architecture to extract the vital features of images and create a latent space representation of an RDF. Subsequently, we use random forest regression to establish correlations between the composition parameters and the latent space representation of an RDF. Finally, the decoder is used to construct an RDF from the latent space variables that are predicted by a random forest for an unknown composition matrix. In principle, any dimensionality reduction technique such as deep neural network (DNN) autoencoders, principal component analysis (PCA), or t-distributed stochastic neighbor embedding (t-SNE) can be used for this purpose. However, the choice of method for dimensionality reduction plays a vital role in the model's overall performance. Our initial studies suggest that convolutional neural network autoencoder-based feature extraction from an RDF is most efficient for the current problem, and it improves the composition-structure ML model of PNCs dramatically. In summary, we develop a robust metamodel that enables intelligent interpolations in the physicochemical space of a polymer nanocomposite without direct simulations or experiments. In the present study, we use a very generic phenomenological model of polymer nanocomposite to construct the machine learning framework that predicts the nanoparticle distribution. This framework can be further expanded to study and predict structures and morphologies of other nanocomposite systems, including polymer-grafted nanoparticles, and polymer-grafted nanoparticles in a polymer matrix. Moreover, all-atom molecular simulations and experimental data can be utilized to build the ML pipeline. We expect that the mathematical framework of the nanoNET will be useful for predicting any spatial or temporal correlation of molecular systems.

**ACKNOWLEDGMENT**

The work is made possible by financial support from the SERB, DST, and Gov. of India through a start-up research grant (SRG/2020/001045) and the National Supercomputing Mission's research grant (DST/NSM/R&D_HPC_Applications/2021/40). This research uses resources of the Argonne Leadership Computing Facility, which is a DOE Office of Science User Facility supported under Contract DE-AC02-06CH11357. We also use the computational facility of the Center for Nanoscience Materials. Use of the Center for Nanoscale Materials, an Office of Science user facility, was supported by the U.S. Department of Energy, Office of Science, Office of Basic Energy Sciences, under Contract No. DE-AC02-06CH11357.We acknowledge the use of the computing resources at HPCE, IIT Madras.